# Strain-driven magnetic anisotropy and spin reorientation in epitaxial $CoV_2O_4$ spinel oxide thin films

Lamiae El Khabchi[a], Laurent Schlur[a], Jérôme Robert[a], Marc Lenertz[a], Cédric Leuvrey[a], Gilles Versini[a], François Roulland[a], Gilbert Chahine[b], Nils Blanc[c], Daniele Preziosi[a], Christophe Lefèvre[a], and Nathalie Viart[a,*]

[a] Université de Strasbourg, CNRS, IPCMS, UMR 7504, 23 rue du Loess, 67034 Strasbourg, France
[b] SIMaP, Université Grenoble Alpes, CNRS, Grenoble INP, 38000 Grenoble, France
[c] Université Grenoble Alpes, CNRS, Grenoble INP, Institut Néel, 38000, Grenoble, France

* Corresponding author: nathalie.viart@ipcms.unistra.fr

**ABSTRACT**

$CoV_2O_4$ (CVO) stands out among spinel vanadates for its ultra-short V–V distances, placing it at the brink of itinerant electron behaviour—an ideal playground for strain engineering. In this work, we exploit this sensitivity by growing high-quality epitaxial CVO thin films on $SrTiO_3$ (001) and MgO (001), inducing compressive and tensile strain, respectively. Using pulsed laser deposition under ultra-low oxygen pressure, we achieve high crystalline quality and strain-controlled tetragonal distortions: $c > a$ under compression (STO) and $c < a$ under tension (MgO). Resonant elastic X-ray scattering confirms a normal spinel structure, with cobalt occupying tetrahedral sites and vanadium octahedral ones. Both strain types reduce charge transport, driving the system into a highly resistive state. Magnetic measurements reveal strain-driven anisotropy switching: STO films transition from out-of-plane to in-plane easy axis below 90 K, while MgO films flip from in-plane to out-of-plane below 45 K. These results highlight CVO's exceptional responsiveness to lattice strain, unlocking a path to finely tunable electronic and magnetic properties. With its strong spin-lattice coupling and potential in spin Hall magnetoresistance, strained CVO emerges as a compelling platform for next-generation low-power spintronic devices.

# I. INTRODUCTION

The complex interaction between lattice dynamics and electronic degrees of freedom (charge, spin and orbital) in transition metal oxides gives rise to fascinating physical phenomena and new electronic phases. In such systems, even subtle perturbations can significantly alter the properties, potentially leading to the emergence of new states of matter [1,2]. Among the various classes of transition metal oxides, spinel vanadates $AV_2O_4$ (A being a transition metal) stand out due to their complex behaviours which can be controlled through the manipulation of the distance between the vanadium atoms V-V ($d_{V-V}$) [3–6]. These materials exhibit geometrical frustration, orbital order, electron itinerancy and competing magnetic interactions, resulting in intricate magnetic structures at low temperatures. When non-magnetic atoms are on the A-sites (A=Zn, Cd, Mg), the material undergoes two successive transitions with decreasing temperature [7–9]. The first one, structural, from cubic to tetragonal symmetry, relieves the orbital degeneracy of the $V^{3+}$ $t_{2g}^2$ orbitals, and results in orbital ordering. The second one, magnetic, witnesses the establishment of an antiferromagnetic order. Spinel vanadates with a magnetic ion at the A-site (A = Fe, Mn), display additional magnetic exchange interactions $J_{A^{2+}-A^{2+}}$ and $J_{A^{2+}-V^{3+}}$, which come to play alongside with the $J_{V^{3+}-V^{3+}}$. The interplay of these competing exchange interactions with spin-orbit coupling and Jahn-Teller distortions on the V-site results in multiple structural phase transitions, culminating in low-temperature non-collinear orbitally ordered ground states [5,10,11].

In this context, $CoV_2O_4$ (CVO) emerges as a material of significant interest because it shows the shortest V-V distances of all spinel vanadates ($d_{V-V}$= 2.9724 Å) [12], which places it on the brink of an itinerant-localized crossover regime [12–14]. It was first synthesized in its bulk form in the early 1960s, and proved to be fascinating for both its collective-to-localized-electron behaviour transition and magnetic collinear-to-non-collinear instabilities [15,16]. Bulk CVO adopts a normal spinel cubic structure with a $Fd\bar{3}m$ space group and a cell parameter of $a$=8.407 Å [ICDD PDF file #04-002-4958] or $a$=8.41160(1) Å [17]. Some doubts on the normal character of the $CoV_2O_4$ spinel have been emitted [15], but one usually admits that $V^{3+}$ cations stay on the octahedral sites due to their relative stabilization in an octahedral crystal field [18], and this has been experimentally checked by neutron diffraction Rietveld refinements [17]. The situation is then that magnetic $Co^{2+}$ ions ($3d^7$, $(e)^4(t_2)^3$, S = 3/2) sit on the A-sites, and magnetic and orbitally active $V^{3+}$ ions ($3d^2$, $(t_{2g})^2(e_g)^0$, S=1) sit on the B-sites. In its bulk form, CVO shows three magnetic transitions and two structural ones, at temperatures slightly varying in the literature, most likely because of stoichiometry variations [17]. Upon cooling, CVO first undergoes a transition from a paramagnetic (PM) to a collinear ferrimagnetic (CL FIM) state at $T_1$ ~ 142 K while keeping its $Fd\bar{3}m$ cubic structure. This is followed by a CL FIM to non-collinear ferrimagnetic (NC FIM) transition at T* ~ 95 K, with a canting of the spins of the V atoms of about 11°, accompanied by a structural transition from cubic $Fd\bar{3}m$ to tetragonal $I4_1/amd$. Finally, at $T_2$ ~ 59 K, one observes an antiferro-orbital ordering (AF-OO) of the $V^{3+}$ ions, along with a structural change from $I4_1/amd$ to $I4_1/a$ [17]. CVO is thought to stand out due to its strongly itinerant character, which reduces its need for orbital degeneracy relief. Its cubic to tetragonal transition is very subtle ((1 − c/a) < 0.06 %) and was only recently experimentally uncovered [17,19], overturning the long-held belief that, unlike other spinel vanadates $AV_2O_4$, CVO remained cubic down to the lowest temperatures [12].

The strong interplay between spin frustration and orbital degrees of freedom exacerbated by short V-V distances offers unique possibilities to manipulate both conductivity and magnetism through a simple external perturbation such as pressure. A possible method for applying pressure is through substrate-induced stress when the material is deposited as a thin film. This has sparked recent interest in the literature regarding the deposition of CVO in thin films, either by pulsed laser deposition or radio frequency sputtering, however with still only few reports until now [20–25]. Films deposited under compressive or tensile stress indeed exhibit markedly different behaviours in their magnetic anisotropy. Similarly to what is observed in bulk, in both cases, two magnetic transitions are observed: one around 150 K, marking the onset of magnetic order, and another one at lower temperature, associated with a major spin reorientation, but the magnetic anisotropy is first out-of-plane and then in-plane for a compressive stress [20–22] and first in-plane and then out-of-plane for a tensile one [25]. Although all works suggest that these different magnetic behaviours originate from structural changes induced by substrate-imposed stress, only two of them present a detailed crystallographic study, and diverge on the exact symmetry adopted by the CVO films, which is orthorhombic for Thompson et *al.* [20,22] and tetragonal for Behera et *al.* [21].

These first, and sometimes conflicting, results motivated us to investigate further the structural, magnetic and electrical properties of strained CVO thin films. Depositions on both STO (001) and MgO (001) substrates allowed us to study the effect of both compressive and tensile strain, respectively, for samples elaborated in the same conditions with the same pulsed laser deposition set-up. We chose to optimize the deposition of the films using an original approach based on the use of a stoichiometric $CoV_2O_4$ target. Until now, the targets used for the deposition of CVO thin films have either been metal Co and V targets, in the case of radio frequency magnetron sputtering [24] or a $CoV_2O_6$ target, in the case of pulsed laser deposition (PLD) [20,21,23,25], but never a stoichiometric $CoV_2O_4$ target. Given the primary importance of oxygen content in oxides, we consequently explored the use of this alternative target, and have placed particular emphasis on the structural characterization of the films we deposited. Our results allow us to unambiguously show that CVO always adopts a tetragonal structure, both on STO and MgO, and not an orthorhombic one, with c>(a=b) for STO and c<(a=b) for MgO. The cationic distribution in the films was determined by synchrotron radiation-based resonant elastic scattering experiments, which confirmed the normal character of the spinel structure of CVO films. We observed opposite behaviour of the magnetic anisotropy for opposite substrate-induced stresses, with clear magnetic transitions for both cases. The spin re-orientation occurs from out-of-plane to in-plane at ca. 90 K (in decreasing temperatures) for films deposited under a compressive strain (STO), while the magnetization easy axis switches from in-plane to out-of-plane at ca. 45 K for films deposited under a tensile strain (MgO). The ability to generate and control distinct magnetic anisotropy states, enabled by the high crystalline quality of the films, represents a valuable asset for a wide range of magnetic applications, particularly in the development of next-generation spintronic devices. Spinel vanadium-based oxides have already proven to be of interest for spin Hall magnetoresistance-based systems and low-power spintronics [26]. This work broadens the currently envisaged perspectives by introducing tuneable magnetic anisotropy in the Pt/CVO//STO|MgO heterostructures.

## II. MATERIALS AND METHODS

Ca. 45 nm thick $CoV_2O_4$ thin films were grown by Pulsed Laser Deposition (PLD) using a krypton fluoride (KrF) excimer laser (COHERENT COMPex Pro 102, λ=248 nm) to ablate a stoichiometric spinel $CoV_2O_4$ single-phased target. The synthesis of this target by an optimized solid-state ceramic method has been described in a previous work [27]. The films were grown on two different substrates, (001) $SrTiO_3$ (STO) (*a*=3.905 Å, CODEX International) and (001) MgO (*a*=4.211 Å, CODEX International) single crystals (0.5 mm thick), yielding opposite strains through the lattice mismatch with CVO, compressive ($\Delta_{CVO/STO}= -7.6\%$) and tensile ($\Delta_{CVO/MgO}= +0.2\%$), respectively. We define here the lattice mismatch as $\Delta = \frac{2 \times a_{substrate} - a_{CVO\ bulk}}{2 \times a_{substrate}}$. Both substrates were cut into $5 \times 5\ mm^2$ pieces. Their surfaces were cleaned in ultrasonic baths successively with acetone, ethanol and isopropanol (10 min for each solvent). The STO substrates were not further treated to reach single termination. The distance between the target and the substrate in the PLD set-up is of 55 mm. The 60 mJ and $0.95 \times 1.35$ mm² laser spot is scanned over an area of $3 \times 3$ mm², resulting in a fluence of 4.68 J/cm². A pre-ablation of the target (3000 laser shots with a repetition rate of 10 Hz and under an oxygen pressure of $3 \times 10^{-5}$ mbar) was performed prior to any film deposition. The films were then deposited with a repetition rate of 5 Hz under an oxygen pressure of $3 \times 10^{-5}$ mbar. The base pressure of the PLD chamber was in the $10^{-8}$ mbar range. The highest degree of crystallization was obtained for a deposition temperature of 600°C, which was indeed used for depositions onto STO substrates. MgO substrates, which have a crystalline structure very close to the spinel one, showed important interfacial reactivity with the film and the temperature had to be lowered to 500°C to yield qualitative interfaces. After deposition, the films were cooled down to 100 °C at a speed of 15°C/min under the oxygen deposition atmosphere. Reflection high-energy electron diffraction (RHEED) was used to check the quality of the substrates surface prior to deposition and to monitor the growth of the films. The used azimuth was the [100] direction, for both substrates. The topography of the substrates and deposited thin films was characterized using atomic force microscopy (AFM) in the non-contact mode (Park Systems XE7). The composition of the films was checked by energy dispersive spectroscopy (EDS) coupled to a scanning electron microscope (Zeiss GeminiSEM 500). The structural characterization of the films was performed by X-ray diffraction (XRD) in parallel beam geometry using a Rigaku SmartLab diffractometer equipped with a 9 kW rotating copper anode ($K_{\alpha 1}$= 0.154056 nm) and a double bounce Ge (220)x2 front monochromator. The crystallinity of the films and their epitaxial relationship with the substrates were studied in $\theta - 2\theta$ and $\varphi$-scans modes. Reciprocal space mappings (RSM) complemented these measurements and allowed to determine the complete set of lattice parameters of the deposited thin films. The same set-up was used to determine the thickness of the films and the quality of the interfaces by X-ray reflectivity.

Deeper insights into the structure of the films were gained with Resonant Elastic X-ray Scattering (REXS) experiments performed on the collaborating research group (CRG) D2AM beamline at the European Synchrotron Radiation Facility (ESRF, Grenoble) [28]. The cationic distribution and the position of the oxygens were assessed from REXS experiments in the X-ray absorption near edge structure (XANES) domain, also known as diffraction anomalous near edge structure (DANES), and in the extended X-ray absorption fine structure (EXAFS) domain, also known as extended diffraction absorption fine structure (EDAFS), respectively. The REXS spectra were acquired at the Co edge (7.7 keV), spanning by 50 eV below and above the edges, for about a dozen reflections. The samples were mounted on a seven-circle diffractometer and the intensities of the incident beam, the diffraction peaks and the fluorescence of the samples were recorded. The rotation matrix was first determined with the help of different in-plane and out-of-plane reflections. The anomalous factors f' and f'' were extracted from the fluorescence data. The knowledge of $f_0$, f' and f'' allows the simulation of the scans for a given set of structural parameters. The refinement of the experimental spectra was performed by minimizing the sum of all reliability factors for all reflections and edges through a stochastic basin-hopping algorithm [29] using a procedure and software (FitREXS) we have developed and fully described in previous works [30,31].

The magnetic properties of the CVO thin films were characterized using a superconducting quantum interference device magnetometer (SQUID VSM MPMS 5, Quantum Design). Measurements of magnetization as a function of temperature both in zero field cooled (ZFC) and field cooled (FC) modes were performed in the temperature range between 2 and 300 K under a magnetic field of $\mu_0 H$ = 0.05 T for both systems (CVO//STO and CVO//MgO) and in both in-plane (IP) and out-of-plane (OOP) configurations. M vs. H hysteresis loops were also measured to determine the saturation magnetization and coercive fields for various temperatures. The electrical properties of the grown CVO thin films were evaluated from the temperature dependence of their resistance measured in a Van der Pauw configuration using a physical property measurement system (PPMS Dynacool, Quantum Design).

### III. RESULTS AND DISCUSSION

Figure 1 shows the AFM and RHEED images of the deposited films, on both substrates, together with the images of the substrates prior to deposition, for comparison. The high quality of the substrates surface is attested by the rod patterns and Kikuchi lines observed on their RHEED patterns [32]. While the RHEED patterns of the CVO films grown on STO show spots, indicating a 3D-growth, those of the films grown on MgO show streaky patterns, indicating a quasi 2D-growth. The initial roughness (root mean square, rms) of the substrates is slightly lower for STO (ca. 0.4 nm) than for MgO (ca. 0.7 nm). After deposition, however, the CVO films deposited on MgO still have a 0.7 nm roughness while those deposited on STO have an increased roughness of 1.0 nm, probably because the larger lattice mismatch between STO and CVO is less favourable to a smooth layer-by-layer growth. The thicknesses of the $CoV_2O_4$ thin films were determined to be $t_{CVO/STO}$= (59.0 ± 1.0) nm and $t_{CVO/MgO}$= (41.6 ± 0.2) nm from X-ray reflectivity measurements (the fits are shown in Figure S5).

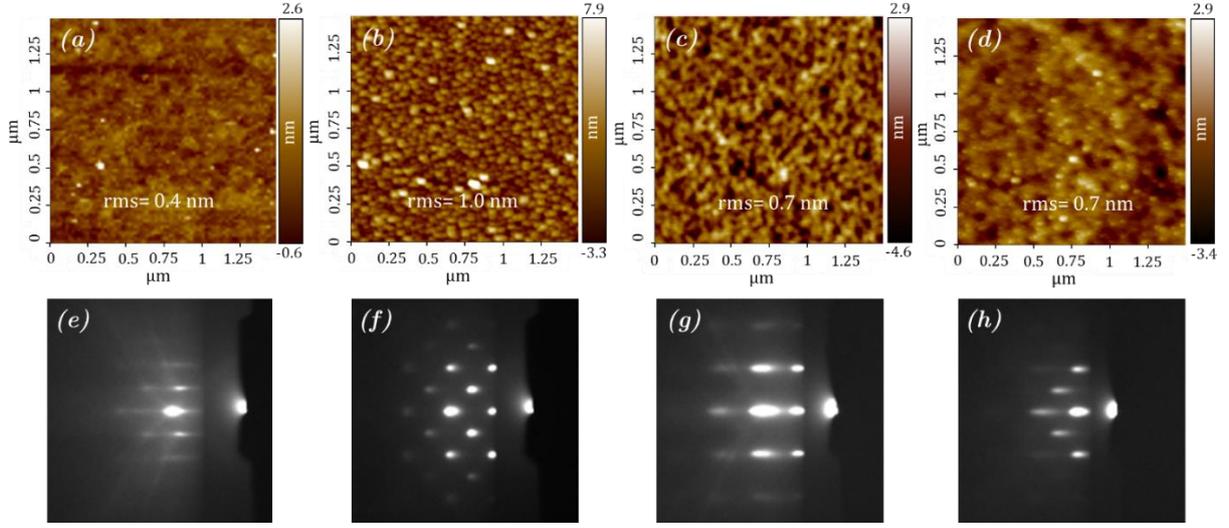

FIG. 1. (a-d) AFM and (e-h) RHEED images of the STO (a,e) and MgO (c,g) substrates prior to deposition and of the CVO deposited films post deposition on STO (b,f) and on MgO (d,h). Root mean square roughness values are indicated on the AFM images.

X-ray diffraction patterns of the deposited films, measured in the $\theta - 2\theta$ mode, are shown in Fig. 2 (a) and (b). They indicate that, for both substrates, the CVO films are well crystallized, oriented along the [001] axis, without any trace of spurious phase. Zooms around the 004 and 008 reflections of the films allow to better seize their relative position with respect to the out-of-plane reflections of the substrates. The presence of some Laue oscillations for the films grown on MgO confirm their high crystallinity and low roughness [33]. The $\varphi$-scans performed for the 013 and 024 reflections of the STO and MgO substrates, respectively, and the $026_c$ reflection of the CVO film are shown in Fig. 2 (c) and (d). The cubic notation of the expected spinel structure is used here for CVO, denoted by 'c', even if the in-plane parameters happen to be different from the out of plane one. The $\varphi$-scans indicate a cube-on-cube type epitaxial growths in both cases, with epitaxial relationships [100] CVO (001) // [100] STO or MgO (001). Figure 3 shows the reciprocal space maps obtained for the 026 reflection of the CVO film grown on the two different substrates. The maps were performed for the four $\varphi$ angles at which such a reflection may be found, thus scanning the $(026)_c$, $(206)_c$, $(0\bar{2}6)_c$ and $(\bar{2}06)_c$ planes. For each substrate, the four maps all show a maximum of intensity at the exact same position in q, yielding unique values for *a* and *b*, and a different one for *c*. This not only certifies the quality of the sample alignment but also provides unquestionable proof of the tetragonal character of the films. The cell parameters calculated from the RSM are $a_{c\ CVO//STO} = b_{c\ CVO//STO} = 0.8362(8)$ nm, $c_{c\ CVO//STO} = 0.8450(4)$ nm for CVO films grown on STO, and $a_{c\ CVO//MgO} = b_{c\ CVO//MgO} = 0.8428(5)$ nm, $c_{c\ CVO//MgO} = 0.8367(2)$ nm for CVO films grown on MgO. These values confirm that the CVO films deposited onto STO are subjected to an in-plane compressive strain, with $a_{c\ CVO//STO} = b_{c\ CVO//STO} < a_{bulk\ CVO}$, while those deposited onto MgO are subjected to a tensile one, with $a_{c\ CVO//MgO} = b_{c\ CVO//MgO} > a_{bulk\ CVO}$. Figure 4 presents the RSM plots that simultaneously show the CVO films $0\bar{2}6_c$ and $0\bar{4}8_c$ reflections together with the substrates $0\bar{1}3$ and $0\bar{2}4$ reflections, for STO and MgO, respectively. One can see that while the CVO films deposited on MgO are fully strained, with a perfect adjustment of their in-plane lattice parameter to that of MgO, those deposited on STO are only partially-strained. This can be explained by the very small lattice mismatch of CVO with respect to MgO $\Delta_{CVO/MgO} = +0.2\%$ compared to the one with STO $\Delta_{CVO/STO} = -7.6\%$. One can also notice that for CVO//STO (CVO//MgO) the in-plane compression (expansion) is accompanied with an out-of-plane expansion (compression). Such behaviours can be quantified by calculating the Poisson ratio ν, which reflects the resistance of a material to deform under a mechanical stress rather than to change its volume [34–36]. For an epitaxial film under a 2D stress, the Poisson ratio (ν) is related to the apparent Poisson ratio (ν*) given by $\nu^* = -\frac{\varepsilon_{OOP}}{\varepsilon_{IP}} = \frac{2\nu}{(1-\nu)}$ [34]; the strains $\varepsilon_{OOP}$ and $\varepsilon_{IP}$ being calculated from the in-plane (IP) and out-of-plane (OOP) lattice parameters $a$ and $c$ of the films as well as the bulk CVO lattice parameter ($a_{bulk} = c_{bulk} = a_{CVO\ bulk} = 8.407$ Å [12]) as $\varepsilon_{OOP} = \frac{(c - c_{bulk})}{c_{bulk}}$ and $\varepsilon_{IP} = \frac{(a - a_{bulk})}{a_{bulk}}$, respectively. Values of $\nu_{CVO/STO} = 0.32$ and $\nu_{CVO/MgO} = 0.47$ have been found. These values are positive, which indicates that the films have a non-auxetic behaviour, i.e. that they expand (compress) OOP when submitted to a compression (an extension) IP, with a view to try and keep their unit cell volume constant. For the growth on MgO the Poisson ratio of 0.47, very close to 0.5, indicates that upon the tensile strain imposed by MgO, the volume of CVO is very close to being conserved. This not the case for the growth on STO, for which the Poisson ratio of 0.32 indicates that under the in-plane compressive strain, there is an out-of-plane elongation (ν>0), but not as much as required to conserve the volume (ν<0.5) (See S.I. for additional information [70]). As a matter of comparison, if some negative values (auxeticity) have been observed for $CoFe_2O_4$ [37–39] and attributed to a peculiar cationic distribution [37,39], one must note that the Poisson's ratios observed for oxides in thin films are majorly positive and commonly within a 0.2-0.4 range [34,40–43]. The Poisson's ratio observed for the CVO//STO thin films is thus fully within the standard range and that of 0.47 observed for the CVO//MgO film is rather at the upper end of what is observed in the literature.

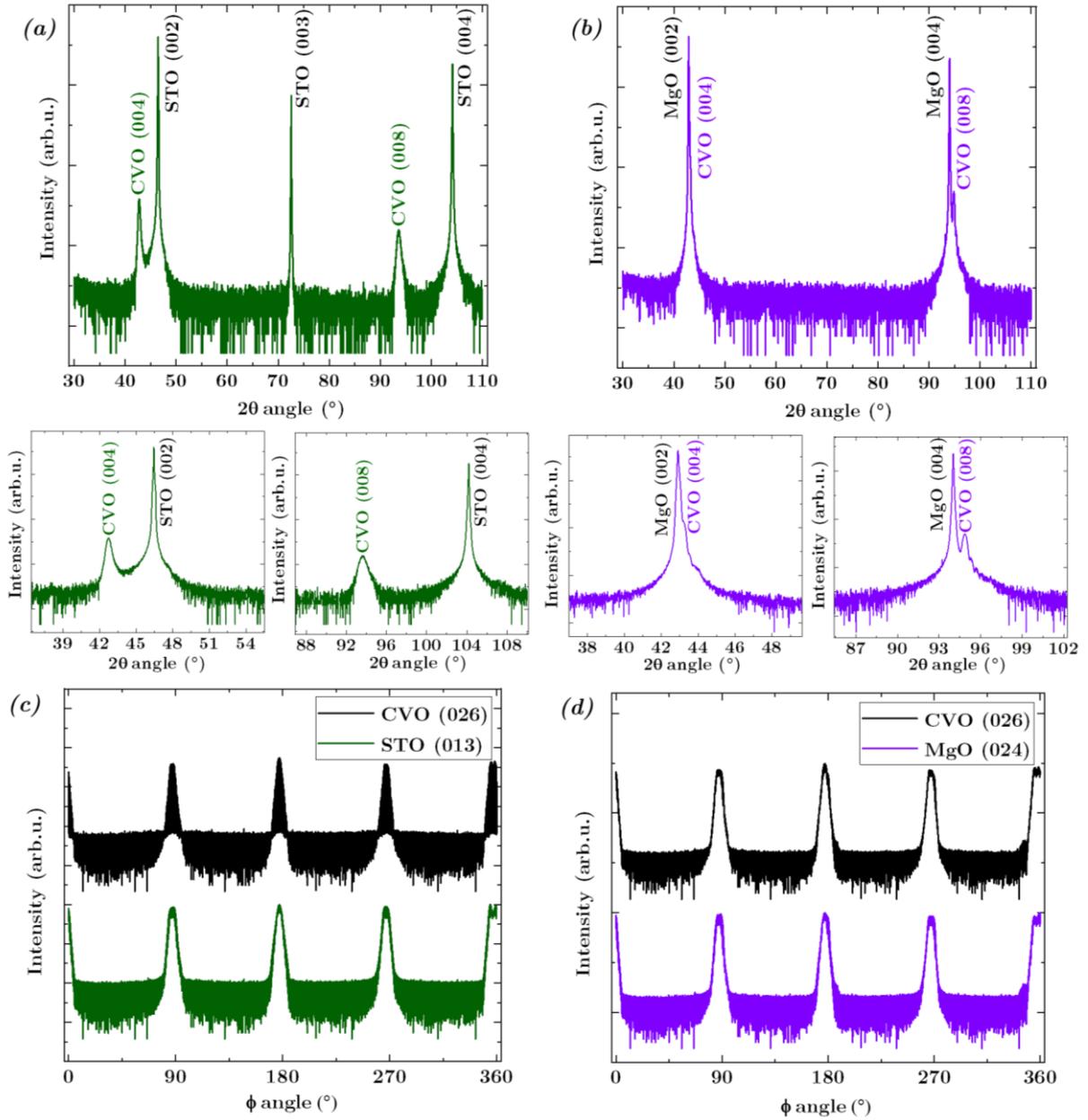

FIG. 2. (a-b) X-ray diffraction patterns in the $\theta - 2\theta$ symmetrical mode of the CVO films on (001)-oriented (a) STO and (b) MgO substrates, with zooms around the films' Bragg reflections, (c-d) $\varphi$-scans of the CVO $\{026\}_c$ alongside with those of the (c) STO $\{013\}$ and (d) MgO $\{024\}$ substrates.

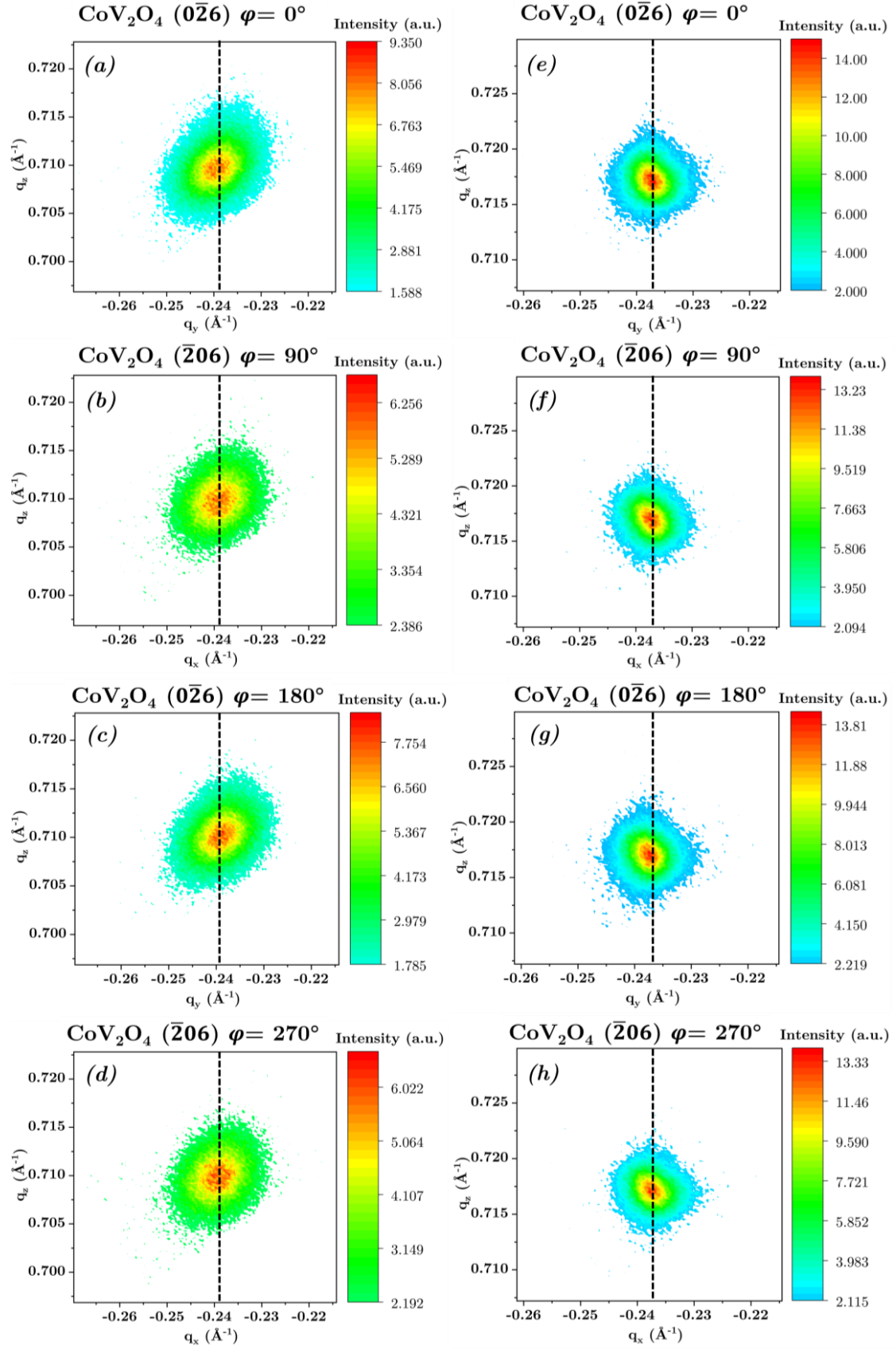

FIG. 3. 2D-reciprocal space mappings for the CVO {026}$_c$ at the four 90°-separated φ angles for depositions on STO (a-d) and MgO (e-h). The vertical black dashed line serves as a guide to the eyes evidencing that all in-plane lattice parameters are identical (a=b) for both substrates.

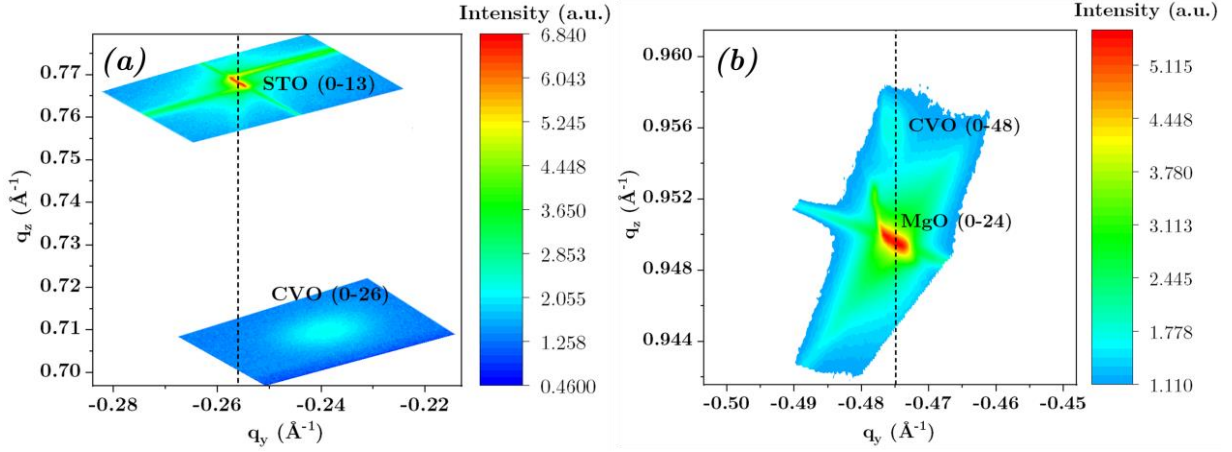

FIG. 4. Reciprocal space mappings of the $0\bar{2}6_c$ and $0\bar{4}8_c$ reflections of the CVO films together with the closest reflections of the (a) STO and (b) MgO substrate.

To gain a deeper knowledge of the crystallographic structure of our CVO thin films, we performed synchrotron light-based resonant elastic X-ray scattering experiments at the Co-edge. Reciprocal space maps were acquired for about a dozen of reflections for each sample (Figures S9 and S10). The experimental angles (θ, ω, φ) were converted into reciprocal space coordinates through projections in the xy, xz, and yz scattering planes. Since the out-of-plane and in-plane lattice parameters are not equal, the conventional $Fd\bar{3}m$ cubic spinel structure is no longer applicable to CVO, and a new crystallographic model, based on a tetragonal symmetry, must be identified prior to any refinement. Using the Cellsub program from the Bilbao Crystallographic Server [44–46], the most plausible tetragonal subgroup derived from $Fd\bar{3}m$ which best preserves the symmetry relations of the original cubic structure was determined to be $I4_1/amd$. The relationships between the cell parameters in the two groups ('c' stands for cubic and 't' for tetragonal) are $a_t = \frac{\sqrt{2}}{2}a_c$, $b_t = \frac{\sqrt{2}}{2}b_c$, and $c_t = c_c$. (Further details on the relationships between the two space groups can be found in the S.I.).

The $q_x$, $q_y$, $q_z$, together with their error bars, were obtained from the numerical determination of the reciprocal-space maps centres. The lattice parameters were then determined by least-squares minimization of the function $f = (E - T)^2$, with $E = q_x^2 + q_y^2 + q_z^2$ and $T = \left(\frac{h}{a}\right)^2 + \left(\frac{k}{a}\right)^2 + \left(\frac{l}{c}\right)^2$. The unit-cell parameters $a$ and $c$ were treated as the only free variables. The entire $q_x$, $q_y$, $q_z$ dataset acquired for one sample was fitted simultaneously. The procedure returns the optimized values of the free parameters, as well as the covariance matrix, whose diagonal elements' square roots correspond to the standard deviations. The thus determined cell parameters, given in the cubic notations to allow a better comparison with the previously determined ones, are $a_{c\ CVO//STO} = b_{c\ CVO//STO} = 0.8361(2)$ nm, $c_{c\ CVO//STO} = 0.84503(6)$ nm for CVO films grown on STO, and $a_{c\ CVO//MgO} = b_{c\ CVO//MgO} = 0.8429(2)$ nm, $c_{c\ CVO//MgO} = 0.8364(2)$ nm for CVO films grown on MgO (Table 1). These values obtained from synchrotron light scattering confirm those obtained by laboratory RSM measurements, while providing higher precision. Energy dependent resonant spectra were acquired for various reflections, among those for which RSM were acquired, in the near-edge (DANES) and extended (EDAFS) domains. The spectra which were too noisy have been discarded and those which were used for the refinements are shown in Figure 5. The refinements were carried out in the $I4_1/amd$ tetragonal space group for both samples, with FitREXS [30,31], which computes the energy-dependent structure factors by combining tabulated nonanomalous Thomson scattering terms $f_0(Q)$ and anomalous terms f'(E) and f''(E) extracted from fluorescence and Kramers–Kronig transformation. The adjustable parameters include cation occupations and oxygen positions. A global cost function, defined as the sum of the reliability factors over all measured reflections and energies, is minimized by a stochastic *basin-hopping* algorithm, followed by local refinement with the L-BFGS-B approach [47], with a numerical convergence factor of 99.9999%. The rigorous determination of error bars on the occupation factor would require evaluating the covariance matrix of the refined parameters, which in turn implies calculating the Hessian of the multivariable cost function. Since FitREXS employs a stochastic basin-hopping procedure combined with local minimizers, the global Hessian is not computed and no formal covariance matrix is provided in the current implementation. Uncertainties on the refined occupancies were however evaluated in one of our previous studies also concerning spinels thin films by a 1000-sample bootstrap approach [48], which yielded standard deviations of ~0.01 on the cationic distribution. Since this procedure is extremely computationally demanding, it was not repeated here; given that the measurement conditions were identical, we conservatively adopt the same uncertainty levels in the present work. The adopted structure can thus be considered a normal spinel structure within a 1 % error bar.

The EXAFS region of the Co K-edge (E > 7730 eV) provides sensitivity to the local environment of the cobalt atoms and EDAFS data thus offers valuable insight into the oxygen positions within the unit cell. For the CVO//MgO film, the recorded signal exhibits well-defined oscillations (Fig. 5(d)), characteristic of EXAFS, which were successfully fitted with FitREXS [30] and allowed determining the oxygen position (Table 1). The signal recorded for the CVO//STO film (Fig. 5(c)) is significantly noisier, probably due to the film/substrate higher mismatch which induces an overall lower crystalline quality of the film and reduces the coherence of the scattered signal, and the position of the oxygen atoms cannot be reliably determined in that case. The position of the 16h oxygen determined to be (0,0.463(2),0.282(2)) for the CVO//MgO sample at room temperature is somehow different from the expected ideal position. The position of oxygen in a non-distorted spinel expressed in the cubic $Fd\bar{3}m$ space group is (u,u,u), with u ca. 0.26 (typical for a cubic spinel with similar cation sizes). It will therefore be (0, 1/4+u, u) in the tetragonal $I4_1/amd$ space group, i.e. (0,0.51,0.26) in order for the tetragonal structure to reproduce the cubic bond lengths and angles for a=b=c. Here, we have a distorted structure with a=b>c. The fact that the in-plane O are slightly shifted inward (y is 0.463 instead of 0.51) while the out-of-plane ones are shifted outward (z is 0.282 instead of 0.26) partially counteracts the in-plane elongation and out-of-plane compression, respectively. This probably helps maintaining the distances in the local polyhedral oxygen environments of the metallic cations as close as possible to the ideal ones. The structure observed here for CVO is that of room temperature, thus corresponding to the paramagnetic state. The same kind of displacements of the oxygen atoms -inward (outward) shift of the in-plane (out-of-plane) oxygens, with an O position at (0,0.4793(3),0.26053(2))- are also observed for the tetragonal structure observed for bulk CVO at 65 K (also $I4_1/amd$, with a=b<c) [17]. At this temperature, the bulk material is in a non colinear magnetic structure, and it is very likely that these displacements can strongly modify the magnetic anisotropy and play a role in the observed non-collinearity of the moments.

TABLE 1. Cell parameters (in the cubic notation) and reduced coordinates of the oxygen atoms in the tetragonal $I4_1/amd$ space group as determined from synchrotron X-ray diffraction experiments.

|  | CVO//STO | CVO//MgO |
|---|---|---|
| $a_c = b_c$ (nm) | 0.8361(2) | 0.8429(2) |
| $c_c$ (nm) | 0.84503(6) | 0.8364(2) |
| O 16h ($I4_1/amd$) y | - | 0.463(2) |
| z | - | 0.282(2) |

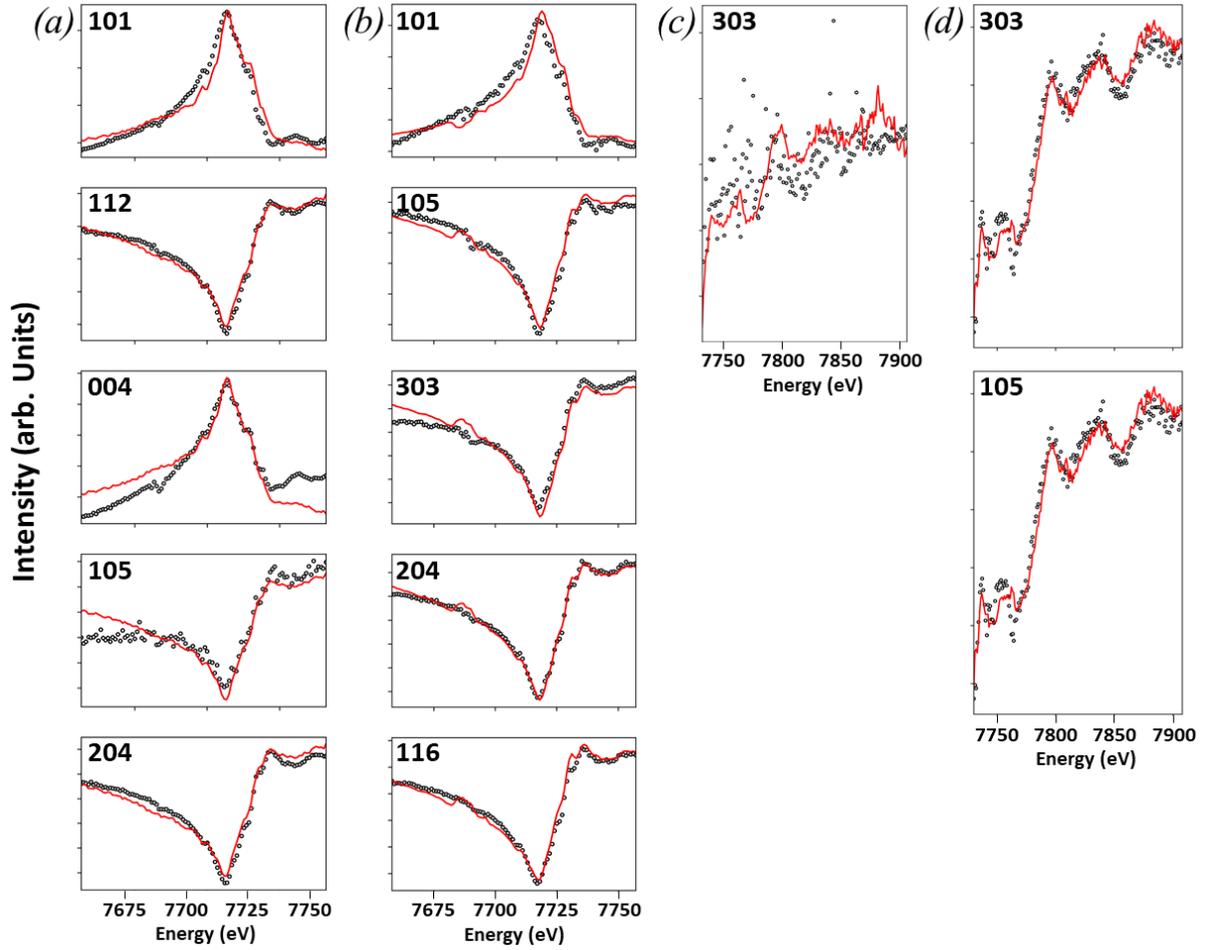

FIG. 5. REXS spectra acquired at room temperature for various reflections at the Co-edge in the near-edge (DANES) (a,b) and extended (EDAFS) (c,d) domains for the CVO films deposited onto STO (a,c) and MgO (b,d) substrates (points are the raw data and red curves are the refinement).

The temperature variation of magnetization in both ZFC and FC modes, with an applied magnetic field of $\mu_0 H = 0.05$ T, is shown in Fig. 6 (a-b) for both systems, in IP (magnetic field applied along the *a* or *b* axis) and OOP (magnetic field applied along the *c* axis) configurations. The two films show two clear magnetic transitions with completely opposite magnetic anisotropies. For CVO//STO (CVO//MgO), magnetization first settles down at 150 K (127 K) in the OOP (IP) direction, and then re-orientates from OOP to IP (IP to OOP) at 90 K (45 K). The magnetic transition temperatures were determined as extrema in the first derivative curves. The first transition coincides well with the one observed for single- and poly-crystalline $CoV_2O_4$ bulk samples [4,12,17,27,49] and thin films [20,21,24,25], and corresponds to the transition from a paramagnetic to a CL FIM state. The second transition corresponds to a transition towards a NC FIM state, as observed in bulk CVO [14,17,19]. Thompson et *al*. [20] indeed evidenced a 20° canting of the V moments with respect to the direction of the Co ones in this NC FIM state in CVO films deposited onto STO by means of neutrons diffraction.

The different magnetic behaviours of CVO thin films under compressive or tensile stress have been explained with the help of density functional theory [25]. If it is simply accounted for by a competition between magnetocrystalline and shape anisotropies in the case of compressive stress, the strong magnetostriction of CVO [23] has to be invoked to account for the magnetic behaviour of CVO thin films under tensile stress. The particularly sharp and distinct magnetic transitions displayed by the ZFC and FC curves result from the high crystalline quality of the films with well-defined structural distortions.

Hysteresis loops measured for the two systems in both parallel and perpendicular configurations are shown in Fig. 6 (c-d). Measurements were conducted at different temperatures, each representative of a distinct magnetic anisotropy state of the films. The slope originating from the diamagnetic signal of the substrates was subtracted from the curves, assuming that the samples magnetization reaches saturation for the highest fields. The hystereses show a saturation magnetization of $\approx 1.5$ $\mu_B$/f.u for the lowest measured temperatures for both samples (40 K for CVO//STO and 30 K for CVO//MgO films), in good agreement with what is observed in the literature for bulk at similar temperatures [4,27]. The usual superexchange interactions in spinels will lead to an anti-ferromagnetic coupling between the spins of the A-sites $Co^{2+}$ ions and those of the B-sites $V^{3+}$ ones. If $Co^{2+}$ and $V^{3+}$ had their expected moments of 3 and 2 $\mu_B$, respectively, the total saturated magnetic moment would be of $2 \times 2 - 3 = 1$ $\mu_B$/f.u. Values lower than 2 $\mu_B$ have however been reported for the moment of the V atoms, even for materials in the bulk form, and one can find values such as 0.85 (at 4 K), 1.3 (at 5 K), and 0.65 (at 1.8 K) $\mu_B$/at for $FeV_2O_4$ [10], $MnV_2O_4$ [50], and $ZnV_2O_4$ [51], respectively, in the literature. The values commonly observed for the moment of $V^{3+}$ in $CoV_2O_4$, as determined for the bulk form by neutron diffraction measurements, vary between 0.47(3) (at 5 K) [14], 0.64(1) (at 12 K) [17] and 0.71 (at 5 K) $\mu_B$/at [52], accounted for either by the strong frustration of the pyrochlore lattice [17,52] or by the increased itinerancy of electrons [14]. A magnetization of ca. 1.5 $\mu_B$/f.u., as observed here, corresponds to a reduced effective value of ca. 0.75 $\mu_B$/V ion, in good agreement thus with what is usually observed for bulk cobalt vanadates, and should not be ascribed to the form of thin films. Lower values of magnetization of ca. $\approx 1$ $\mu_B$/f.u for the higher measured temperatures (120 K for CVO//STO and 90 K for CVO//MgO films), can simply be explained by the thermal agitation which reduces the total magnetization to lower values.

Figure 6(c) confirms that, at 40 K, the magnetic moments in CVO//STO films are more difficult to saturate OOP than IP. The *a* and *b* axes are easy axes and a relatively large coercive field of $\mu_0 H_c \approx 1.41$ T is required to reverse the magnetization in plane. At T ≈ 120 K, the easy axis has visibly switched to the *c*-axis, consistently with the observed anisotropies in the ZFC-FC curves. On the other hand, the CVO//MgO films show a perpendicular magnetic anisotropy for the lowest temperatures (T ≈ 30 K), in which the moments easily reach saturation along the [001] axis. The coercive field is then of $\mu_0 H_c \approx 0.5$ T, larger than the 0.1 T observed in plane. At approximately 90 K, the easy axis is however switched to the *ab* plane, consistent again with the ZFC-FC curves of CVO//MgO films. All these measurements thus confirm the opposite spin-reorientations which happen in both films, at ≈ 90 K for the CVO//STO and ≈ 45 K for the CVO//MgO. One can observe a small step-like jump at zero field in the 40 K IP hysteresis loop of the CVO//STO sample (Fig. 5(c)). Such a step-like jump, which can also be seen as a double switching, has already been reported in the hysteresis loops of a variety of spinels in thin films and addressed in light of the existence of two magnetic phases of different coercivities [26,53–55]. For Rigato *et al.*, who have grown $CoFe_2O_4$ (001) films onto (001) STO substrates, this is due to the presence of a second phase which is a faceted superficial film of pyramidal huts emerging from lower surface energy of the exposed (111) spinel faces with respect to the (001) ones. The relative importance of the step with respect to the overall cycle therefore decreases with increasing thicknesses [54]. Sofin *et al.*, who have studied the magnetization of $Fe_3O_4$ thin films, give another explanation, in terms of antiphase boundary domains (APB). An interfacial antiferromagnetic coupling between APB domains of different sizes explains that it increases the critical field of the small size domains and decreases that of the large size ones, causing the double switching [55]. There is no thickness dependence in this case. Antiphase boundaries are prone to arise in spinel thin films, originating from the existence of equivalent nucleation sites during growth on substrate displaying a higher symmetry (typically what happens during the growth of CVO onto MgO) [56], from the accommodation of the in-plane film-substrate lattice mismatch [57–59] or from the roughness of the substrate [25]. The magnetic coupling is then much disturbed at the APB, with the intra-sublattice superexchange coupling greatly strengthened, while the inter-sublattice superexchange coupling is weakened, and the APB separates oppositely magnetized regions [60]. Finally, considering the significant magnetic frustration present in CVO, another plausible explanation involves some canting of the spins. The observed step, which is a decrease of the magnetization by about 50 %, leading the magnetization to move from 1.5 $\mu_B$/f.u. to about 0.75 $\mu_B$/f.u., cannot be explained by a canting of the V moments, $\mu_V$, alone. Indeed, in the frame of the antiferromagnetically aligned Co and V moments, this would lead to an increase of the magnetization (the moments of Co are less counterbalanced by those of V). Three scenarios are then possible to explain the double step. In the first one, it is the moments of the Co atoms, $\mu_{Co}$, which become canted. To account for the observed decrease in moment the canting need to be of $\mu_{Co} \cos\alpha - 2\mu_V = 0.75~\mu_B$, i.e. the canting angle a needs to be of about 41°. In the second one, both the Co and V moments are simultaneously canted with respect to the applied field, of an angle $\alpha$ for the Co and $\theta$ for the V. The moment lost is then written as $0.75 = \mu_{Co} \cos\alpha - 2\mu_V \cos\theta$. Various solutions are possible, with $\alpha$ increasing from about 41° for $\theta=0°$ to 60° for $\theta=60°$. This means that the Co moments may be even more canted and make a lesser contribution to the global moment if the V moments are also canted. Finally, in a third scenario, the V sublattice may become magnetically disordered due to frustration, giving rise to randomly oriented spin components rather than a uniform canting. The Co moments may also lose their collinearity at the same time, under the influence of the V moments. The net effect is a reduction of the coherent sum of the Co and V magnetic moments. This scenario is consistent with the expected physics of $CoV_2O_4$, where the strong V–V frustration can indeed drive a collective reorganization of the Co–V network [21,25]. Several mechanisms can thus account for the double step observed in the 40 K in-plane hysteresis cycle of our CVO//STO sample, but identifying its exact origin is beyond the scope of the present work.

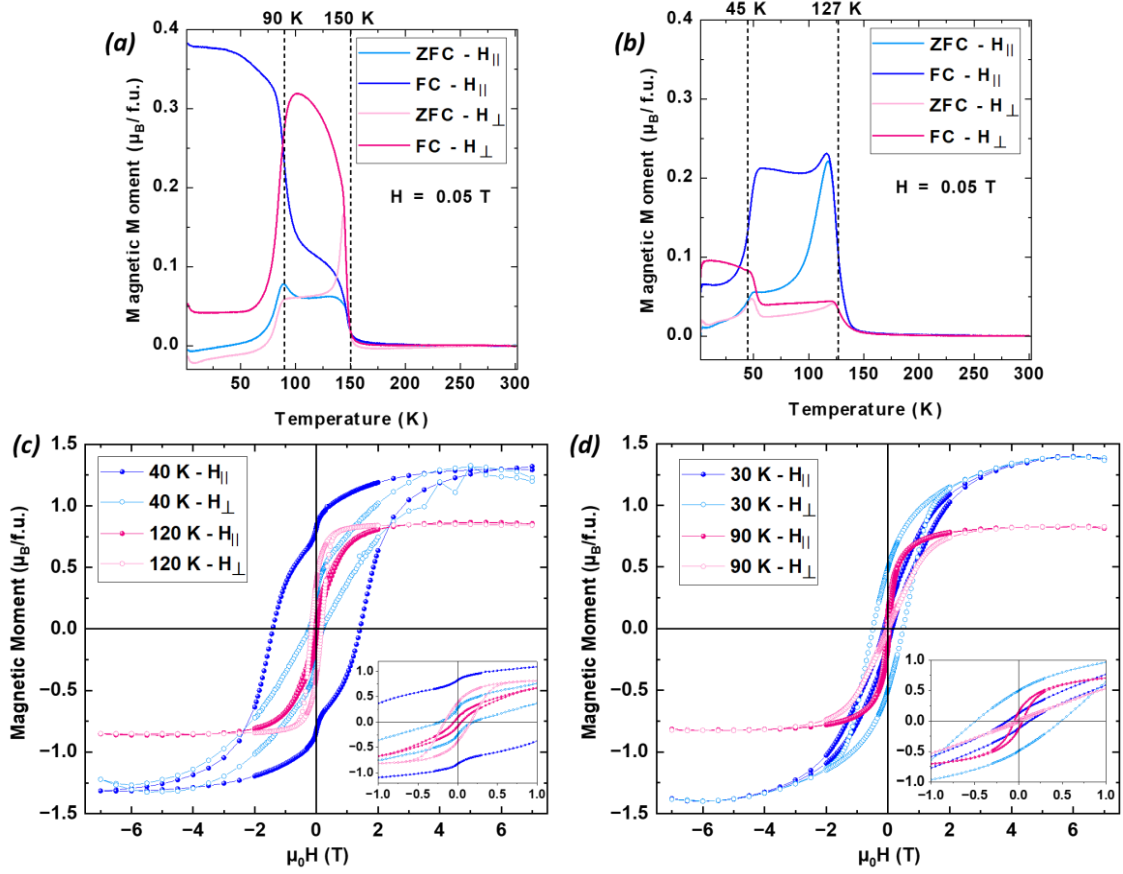

FIG. 6. (a-b) ZFC and FC curves in the IP and OOP configurations of CVO deposited on (a) STO and (b) MgO with an applied magnetic field of $\mu_0 H = 0.05$ T, and (c-d) hysteresis loops measured for CVO thin films at different temperatures in the IP and OOP configurations deposited on (c) STO and (d) MgO substrates, respectively. The hysteresis loops plots show the magnetic moment per formula unit, calculated from the measured magnetization versus applied magnetic field data. When the field is applied IP, it is applied along the [100] direction.

The transport properties of CVO films have been evaluated by temperature dependent resistivity measurements performed using the Van Der Pauw method [61]. Figure 7(a) shows the resistivity (in log scale) of the grown CVO films in the temperature range 100-300 K (for temperatures below 100 K, the resistance values were too high to be measured with our set-up). Both CVO//STO and CVO//MgO films behave as semiconductors, as indicated by their decreasing resistivity with increasing temperature. Both films show lower conductivities than what is observed in bulk since $\rho_{bulk\ CVO} \approx 0.065$ Ω.cm at 300 K [5], however CVO films deposited onto MgO are more conductive than those deposited onto STO, with $\rho_{CVO/MgO} \approx$ 0.206 Ω.cm and $\rho_{CVO/STO} \approx 0.735$ Ω.cm at 300 K (i.e. a resistivity ratio of about $\frac{\rho_{CVO/STO}}{\rho_{CVO/MgO}} \approx$ 3.6). This could be due to the lower level of stress experienced by CVO when deposited on MgO compared to STO, thereby resulting in conduction mechanisms more similar to those found in the bulk material. Our attempts to fit the measured resistivity curves with the Arrhenius model [62,63] were unsuccessful (FIG. S13 in S.I.), because of the presence of strongly localized electronic states. We therefore used the nearest-neighbour hopping (NNH) model, which is the conduction mechanism described in the literature for most of the $AV_2O_4$ vanadates [12,24,64]. In this model, the electrical resistivity varies with temperature according to the following expression:

$$\rho(T) \propto T\, e^{\frac{1}{T}}. \quad (1)$$

Figure 7(b) plots $\ln(\frac{\rho}{T})$ as a function of $\frac{1000}{T}$. The measured resistivities fit with the NNH model, but only for temperatures below the CL FIM to PM state, for both systems. To describe the conduction mechanism at higher temperatures, we used another model, the variable range hopping (VRH), for which resistivity is expressed according to:

$$\rho(T) \propto e^{[(\frac{1}{T})^s]}, \quad (2)$$

where $s$ is the hopping exponent defined by $s = \frac{1}{(d+1)}$ in which $d$ designates the number of conduction dimensions. $s$ can be obtained by plotting $\ln(w(T))$ as a function of $\ln(T)$ (See S.I. for calculation details), where $w(T) = -\frac{\partial \ln(\rho)}{\partial \ln(T)}$ [24,62,65,66]. Figure 7(c) depicts $\ln(w(T))$ as a function of $\ln(T)$. The fits, performed for temperatures above the magnetic order transition, give $d \approx 1$. This suggests that at higher temperatures the conduction mechanism is a one-dimensional VRH one, for CVO films deposited on both substrates. These results are in perfect agreement with what was reported by Hidaka et al. [24] for CVO thin films grown on MgO substrates.

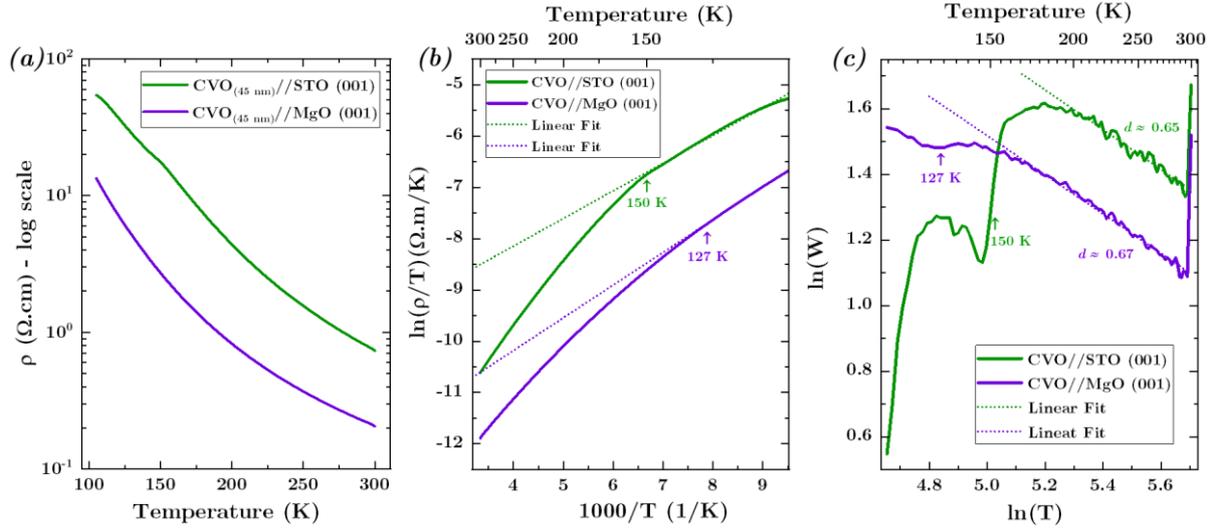

FIG. 7. Electrical properties of CVO films deposited on STO and MgO substrates. (a) Resistivity $\rho$ in log scale as a function of $T$, (b) $\ln(\frac{\rho}{T})$ as a function of $\frac{1000}{T}$, and (c) $\ln(w)$ as a function of $\ln(T)$.

## IV.  CONCLUSIONS

We report here the optimized growth of single-phased CVO thin films on both STO (001) and MgO (001) substrates using the PLD technique. XRD measurements demonstrate that both films are of high crystalline quality and adopt a tetragonal structure with c>(a=b) for STO and c<(a=b) for MgO, due to the compressive and tensile stress imposed by the substrate, respectively. A REXS analysis shows the direct character of the spinel structure adopted by the films deposited on both substrates, with $Co^{2+}$ ions in the tetrahedral sites and $V^{3+}$ in the octahedral ones. The films display a low conductivity in both cases, and their transport properties are dominated by the nearest-neighbour hopping conduction model at temperatures lower than the magnetic ordering temperature and by a one-dimensional variable range hopping conduction mechanism at higher temperatures. They demonstrate two sharp magnetic transitions: the onset of a magnetic order at 150 K (127 K) with an out-of-plane (in-plane) easy magnetization direction, and a spin re-orientation at 90 K (45 K) towards an in-plane (out-of-plane) easy magnetization direction for a growth on STO (MgO). Those completely opposite magnetic behaviours dictated by the opposite substrate-induced strains herald new possibilities of tuneable applications for low power spintronics.

**ACKNOWLEDGEMENTS**

This work was supported by France 2030 government investment plan managed by the French National Research Agency under grant reference PEPR SPIN – SPINMAT ANR-22-EXSP-0007, as well as by the Interdisciplinary Thematic Institute QMat, which as part of the ITI 2021 2028 program of the University of Strasbourg, CNRS and Inserm, was supported by IdEx Unistra (ANR 10 IDEX 0002), and by SFRI STRAT'US project (ANR 20 SFRI 0012) and ANR-11-LABX-0058-NIE and ANR-17-EURE-0024. The authors acknowledge experimental support from the PLD, XRD, MEB-CRO, TEM and MagTransCS platforms of the IPCMS for providing elaboration and characterization facilities. We also acknowledge the European Synchrotron Radiation Facility (ESRF, Grenoble, France) for provision of synchrotron radiation facilities and we would like to thank all the staff for assistance in using the beamline BM02-D2AM.